\documentclass[useAMS,usenatbib]{mn2e}
\usepackage{graphicx}
\newcommand{\bvec}[1]{\mbox{\boldmath $#1$}}

\title[Analytical studies on the SZ effect II]{Analytical studies on the Sunyaev$-$Zeldovich effect in the cluster of galaxies for three Lorentz frames II: single integral formula}
\author[S. Nozawa and Y. Kohyama]{Satoshi Nozawa$^{1}$\thanks{E-mail: snozawa@josai.ac.jp} and Yasuharu Kohyama$^{2}$\\
$^{1}$Josai Junior College, 1-1 Keyakidai, Sakado-shi, Saitama, 350-0295, 
Japan\\
$^{2}$Advanced Simulation Technology of Mechanics Co. Ltd, 2-3-13 Minami, Wako-shi, Saitama, 351-0104, Japan}

\begin{document}

\date{submitted 2013 February **}

\pagerange{\pageref{firstpage}--\pageref{lastpage}} \pubyear{0000}

\maketitle

\label{firstpage}

\begin{abstract}
We study the Sunyaev$-$Zeldovich effect for clusters of galaxies.  The Boltzmann equations for the cosmic microwave background photon distribution function are studied in three Lorentz frames.  We extend the previous work and derive analytic expressions for the integrated photon redistribution functions over the photon frequency.  We also derive analytic expressions in the power series expansion approximation.  By combining two formulas, we offer a simple and accurate tool to analyse observation data.  These formulas are applicable to the non-thermal electron distributions as well as the standard thermal distribution.  The Boltzmann equation is reduced to a single integral form of the electron velocity.
\end{abstract}


\section{Introduction}

  The Sunyaev$-$Zeldovich (SZ) effect \citep{zeld69,suny80}, which arises from the Compton scattering of the cosmic microwave background (CMB) photons by hot electrons in clusters of galaxies (CG), provides a useful method for studies of cosmology.  For the reviews, for example, see \citet{birk99} and \citet{carl02}.  The original SZ formula has been derived from the Kompaneets equation \citep{komp56} in the non-relativistic approximation.  However, X-ray observations, for example, by \citet{alle02} have revealed the existence of high-temperature CG such as $k_{B} T_{e} \simeq $20keV.  For such high-temperature CG, the relativistic corrections will become important.

  On the other hand, it has been known theoretically for some time that the relativistic corrections become significant at the short wavelength region $\lambda < 1$ mm.  In particular, the recent report on the first detection of the SZ effect at $\lambda < 650$ ${\mu}$m by the {\it Herschel} survey \citep{zemc10} seems to confirm the relativistic corrections.  Furthermore, new generation observations, for example, by \citet{plan11} are carrying out systematic studies of the precision measurements on the SZ effect.  Therefore, reliable theoretical studies on the relativistic SZ effect at short wavelength will become extremely important for both existing and forthcoming observation projects.

  The theoretical studies on the relativistic corrections have been done by several groups.  \citet{wrig79} and \citet{reph95} have done the pioneering work to the SZ effect for the CG.  \citet{chal98} and \citet{itoh98} have adopted a relativistically covariant formalism to describe the Compton scattering process and have obtained higher order relativistic corrections to the thermal SZ effect in the form of the Fokker$-$Planck expansion approximation.  \citet{noza98} have extended their method to the case where the CG is moving with a peculiar velocity $\bvec{\beta}_{\rm{c}}$ with respect to the CMB frame and have obtained the relativistic corrections to the kinematical SZ effect.  Their results were confirmed by \citet{chal99} and also by \citet{sazo98}.  \citet{itoh00} have also applied the covariant formalism to the polarization SZ effect \citep{suny81}.  The effect of the motion of the observer was also studied, for example, by \citet{chlu05} and \citet{noza05}.  The importance of the relativistic corrections is also exemplified through the possibility of directly measuring the cluster temperature using purely the SZ effect \citep{hans04}.

  On the other hand, \citet{chlu12} studied the relativistic corrections on the SZ effect by calculating the Boltzmann equation in the CG frame and extending it to other frames.  They reported the importance of the separation of kinetic and scattering terms for the interpretation of future SZ data.  Recently, \cite{noza13} also studied formal relations for the Boltzmann equations among different Lorentz frames, namely for the CMB frame, CG frame and a general observer's (OBS) frame.  They derived analytic expressions for the photon redistribution functions.  With those formulas, the Boltzmann equation for the CMB photon was reduced to two-dimensional integral form.

  In this paper, we extend the work of \cite{noza13}.  We derive analytic expressions for the integrated photon redistribution functions over the photon frequency.  We also derive analytic expressions in the power series expansion approximation.  By combining two formulas, we offer a simple and accurate tool to analyse observation data.  These formulas are applicable to the non-thermal electron distributions as well as the standard thermal distribution.  With these formulas, five-dimensional integrals for solving the Boltzmann equation are reduced to one-dimensional integral of the electron velocity.

  This paper is organized as follows.  In Section 2, we study the Boltzmann equation in the CG frame.  We derive analytic expressions for the integrated photon redistribution functions over the photon frequency.  In Section 3, we derive analytic expressions of the integrated photon redistribution functions in the power series expansion approximation.  In Section 4, we transform the results in the CG frame to the CMB and OBS frames.  In Section 5, we explore numerically the applicable region of the integrated photon distribution functions.  Concluding remarks are given in Section 6.

\section{Calculation in the CG frame}

\subsection{Boltzmann equation}

The kinematics of the CG frame is identical to that of \cite{noza13} except that we dropped the subscript $c$ of the variables in this paper.  The four momenta of the initial and final electrons are $p=(E, \bvec{p})$ and $p^{\prime}=(E^{\prime}, \bvec{p}^{\prime})$, respectively, and $\bvec{\beta}=\bvec{p}/E$ is the velocity of the initial electron and $\gamma=1/\sqrt{1-\beta^{2}}$.  The four momenta of the initial and final CMB photons are $k=(\omega, \bvec{k})$ and $k^{\prime}=(\omega^{\prime},\bvec{k}^{\prime})$, respectively, $x = \omega/k_{B} T_{\rm CMB}$, $x^{\prime} = \omega^{\prime}/k_{B} T_{\rm CMB}$ and ${\rm{e}}^{s} = x^{\prime}/x$.

  We start with equation (26) of \cite{noza13}.  The Boltzmann equation for the CMB photon distribution function $n(x)$ in the CG frame is rewritten as
\begin{eqnarray}
\frac{{\rm{d}} n(x)}{\rm{d} \tau} = \int_{0}^{1} {\rm{d}} \beta \beta^{2} \gamma^{5} p_{e}(\gamma) \, F_{0}(\beta,x) \, - \, n(x)  \nonumber  \\
&& \hspace{-57mm}
+ \, \beta_{\rm{c}} P_{1}(\mu_{\rm{c}}) \left[ \int_{0}^{1} {\rm{d}} \beta \beta^{2} \gamma^{5} p_{e}(\gamma) \, G_{1}(\beta,x) - D \, n(x) \right]  \nonumber  \\
&& \hspace{-57mm}
+ \frac{1}{6} \beta_{\rm{c}}^{2} \left[ \int_{0}^{1} {\rm{d}} \beta \beta^{2} \gamma^{5} p_{e}(\gamma) \Big\{ H_{0}(\beta,x) + 2 \, G_{0}(\beta,x) \Big\} \right. \nonumber  \\
&& \hspace{-13mm}
 -  \, D (D + 2) \, n(x) \bigg]  \nonumber  \\
&& \hspace{-57mm}
+ \, \frac{1}{3} \beta_{\rm{c}}^{2} P_{2}(\mu_{\rm{c}}) \left[ \int_{0}^{1} {\rm{d}} \beta \beta^{2} \gamma^{5} p_{e}(\gamma) \Big\{ H_{2}(\beta,x) \right.  \nonumber  \\
&& \hspace{-34mm}
 - \, G_{2}(\beta,x) \Big\} -  D (D - 1) \, n(x) \bigg]  \, ,
\label{eq2-1-1}
\end{eqnarray}
where $p_{e}(\gamma)$ is the electron distribution function, $\mu_{\rm{c}}$ is the cosine between $\bvec{\beta}_{\rm{c}}$ and $\bvec{k}$, and $D=x \partial/\partial x$.  In equation (\ref{eq2-1-1}), we defined the integrated photon redistribution functions over the photon frequency $s$ as follows:
\begin{eqnarray}
F_{0}(\beta,x) = \int_{-\lambda_{\beta}}^{+\lambda_{\beta}} {\rm{d}}s \, P_{0}(s,\beta) \, n({\rm{e}}^{s}x)   \, ,
\label{eq2-1-2}  \\
&& \hspace{-59mm}
G_{\ell}(\beta,x) = \int_{-\lambda_{\beta}}^{+\lambda_{\beta}} {\rm{d}}s \, P_{\ell}(s,\beta) \, D \, n({\rm{e}}^{s}x)   \, , \, \, \, \ell = 0, 1, 2,
\label{eq2-1-3}  \\
&& \hspace{-59mm}
H_{\ell}(\beta,x) = \int_{-\lambda_{\beta}}^{+\lambda_{\beta}} {\rm{d}}s \, P_{\ell}(s,\beta) \, D^{2} \, n({\rm{e}}^{s}x)  \, , \, \, \, \ell = 0, 1, 2,
\label{eq2-1-4}
\end{eqnarray}
where $\lambda_{\beta}$=ln$[(1+\beta)/(1-\beta)]$, $P_{\ell}(s,\beta)$ are the photon redistribution functions and $n(x)=1/({\rm{e}}^{x}-1)$.  Note that $H_{1}(\beta,x)$ does not appear in equation (\ref{eq2-1-1}), but it does in Section 4.

  One of the main objectives of this paper is to derive analytic expressions for the integrated photon redistribution functions $F_{0}(\beta,x)$, $G_{\ell}(\beta,x)$ and $H_{\ell}(\beta,x)$.  Once they are obtained, five-dimensional integrals for solving the Boltzmann equation (\ref{eq2-1-1}) are reduced to one-dimensional integral of the electron velocity $\beta$.  The obtained formulas are applicable not only to the standard thermal electron distribution but also to arbitrary non-thermal electron distributions.

\subsection{Analytic expressions for the integrated photon redistribution functions}

  In order to derive analytic expressions for $F_{0}(\beta,x)$, $G_{\ell}(\beta,x)$ and $H_{\ell}(\beta,x)$, we show in Appendix A analytic formulas of the photon redistribution functions $P_{\ell}(s,\beta)$.  We also show some useful relations for $P_{\ell}(s,\beta)$ in Appendix A.  

  Inserting equations (\ref{eqA-1}), (\ref{eqA-2}) and (\ref{eqA-28}) into equation (\ref{eq2-1-2}), one obtains the analytic form for $F_{0}(\beta,x)$ as follows:
\begin{eqnarray}
F_{0}(\beta,x) = F_{0,+}(x) - F_{0,+}(\xi x) + F_{0,-}(x) - F_{0,-}(x/\xi) \, ,
\label{eq2-2-1}
\end{eqnarray}
\begin{eqnarray}
F_{0,\pm}(y) =  \alpha \bigg[ \sum_{n=0}^{3} \, a_{n}(\pm \beta) \, \frac{1}{x^{n}} f_{n-1}(y)  \nonumber  \\
&& \hspace{-40mm}
 + \, a_{4}(\pm \beta) \Big\{ \left( \lambda_{\pm \beta} + \ln x \right) \left( \frac{1}{x} f_{0}(y) + \frac{1}{x^{2}} f_{1}(y) \right)  \nonumber  \\
&& \hspace{-24mm}
- \frac{1}{x} f_{0,\ln}(y) - \frac{1}{x^{2}} f_{1,\ln}(y) \Big\} \bigg] \, ,
\label{eq2-2-2}
\end{eqnarray}
where $\xi = (1+\beta)/(1-\beta)$ and $\alpha=3/(32\beta^{2} \gamma^{2})$.  In equation (\ref{eq2-2-2}), we introduced
\begin{eqnarray}
f_{n}(y) = \int_{y}^{\infty} {\rm{d}}u \, \frac{u^{n}}{{\rm{e}}^{u}-1}  \,  , \, \, \,  n=-2, -1, 0, 1, 2, 3,
\label{eq2-2-3}  \\
&& \hspace{-73mm}
f_{n,\ln}(y) = \int_{y}^{\infty} {\rm{d}}u \, \frac{u^{n} \ln u}{{\rm{e}}^{u}-1}  \, , \, \, \, n=0, 1, 2.
\label{eq2-2-4}
\end{eqnarray}
We discuss the analytic structure of the functions $f_{n}(y)$ and $f_{n,\ln}(y)$ in the next subsection.

  The analytic expression for $G_{\ell}(\beta,x)$ is obtained in a similar manner.  First, we apply an integration by parts to equation (\ref{eq2-1-3}) with the relation $D=x\partial/\partial x$=$\partial/\partial s$.  The definite integral term becomes zero with equations (\ref{eqA-29}) and (\ref{eqA-30}).  One obtains
\begin{eqnarray}
G_{\ell}(\beta,x) = - \, G_{\ell,+}(x) + G_{\ell,+}(\xi x) - G_{\ell,-}(x) + G_{\ell,-}(x/\xi)  \, ,
\label{eq2-2-5}
\end{eqnarray}
\begin{eqnarray}
G_{0,\pm}(y) =  \alpha \bigg[ \sum_{n=0}^{3} \, a_{n}(\pm \beta) \, \frac{n}{x^{n}} f_{n-1}(y)  \nonumber  \\
&& \hspace{-40mm}
 + \, a_{4}(\pm \beta) \Big\{ \left( \lambda_{\pm \beta} + \ln x \right) \left( \frac{1}{x} f_{0}(y) + \frac{2}{x^{2}} f_{1}(y) \right)  \nonumber  \\
&& \hspace{-24mm}
 - \frac{1}{x} f_{0,\ln}(y) - \frac{2}{x^{2}} f_{1,\ln}(y)  \nonumber  \\
&& \hspace{-24mm}
 - \frac{1}{x} f_{0}(y) - \frac{1}{x^{2}} f_{1}(y)  \Big\}  \bigg] \, ,
\label{eq2-2-6}
\end{eqnarray}
\begin{eqnarray}
G_{1,\pm}(y) =  \alpha \bigg[ \sum_{n=0}^{3} \, b_{n}(\pm \beta) \, \frac{n}{x^{n}} f_{n-1}(y)  \nonumber  \\
&& \hspace{-40mm}
 + \, b_{4}(\pm \beta) \Big\{ \left( \lambda_{\pm \beta} + \ln x \right) \frac{3}{x^{3}} f_{2}(y)  \nonumber  \\
&& \hspace{-24mm}
- \frac{3}{x^{3}} f_{2,\ln}(y) - f_{-1}(y) - \frac{1}{x^{3}} f_{2}(y)  \Big\}  \nonumber  \\
&& \hspace{-40mm}
 + \, b_{5}(\pm \beta) \Big\{ \left( \lambda_{\pm \beta} + \ln x \right) \left( \frac{1}{x} f_{0}(y) + \frac{2}{x^{2}} f_{1}(y) \right)  \nonumber  \\
&& \hspace{-24mm}
 - \frac{1}{x} f_{0,\ln}(y) - \frac{2}{x^{2}} f_{1,\ln}(y) \nonumber  \\
&& \hspace{-24mm}
 - \frac{1}{x} f_{0}(y) - \frac{1}{x^{2}} f_{1}(y)  \Big\}  \bigg] \, ,
\label{eq2-2-7}
\end{eqnarray}
\begin{eqnarray}
G_{2,\pm}(y) =  \alpha \bigg[ \sum_{n=-1}^{4} \, c_{n}(\pm \beta) \, \frac{n}{x^{n}} f_{n-1}(y)  \nonumber  \\
&& \hspace{-40mm}
 + \, c_{5}(\pm \beta) \Big\{ \left( \lambda_{\pm \beta} + \ln x \right) \frac{3}{x^{3}} f_{2}(y)  \nonumber  \\
&& \hspace{-24mm}
- \frac{3}{x^{3}} f_{2,\ln}(y) - f_{-1}(y) - \frac{1}{x^{3}} f_{2}(y)  \Big\}  \nonumber  \\
&& \hspace{-40mm}
 + \, c_{6}(\pm \beta) \Big\{ \left( \lambda_{\pm \beta} + \ln x \right) \left( \frac{1}{x} f_{0}(y) + \frac{2}{x^{2}} f_{1}(y) \right)  \nonumber  \\
&& \hspace{-24mm}
 - \frac{1}{x} f_{0,\ln}(y) - \frac{2}{x^{2}} f_{1,\ln}(y) \nonumber  \\
&& \hspace{-24mm}
 - \frac{1}{x} f_{0}(y) - \frac{1}{x^{2}} f_{1}(y)  \Big\}  \bigg] \, .
\label{eq2-2-8}
\end{eqnarray}

  Finally, the analytic expression for $H_{\ell}(\beta,x)$ is obtained by applying the integration by parts twice to equation (\ref{eq2-1-4}) with the relation $D^{2}=\partial^{2}/\partial s^{2}$.  One has
\begin{eqnarray}
H_{\ell}(\beta,x) = H_{\ell,+}(x) - H_{\ell,+}(\xi x) + H_{\ell,-}(x) - H_{\ell,-}(x/\xi)  \nonumber  \\
&& \hspace{-71mm}
+ \, H_{\ell,\rm{def}}(x)  \, ,
\label{eq2-2-9}
\end{eqnarray}
\begin{eqnarray}
H_{0,\pm}(y) =  \alpha \bigg[ \sum_{n=0}^{3} \, a_{n}(\pm \beta) \, \frac{n^{2}}{x^{n}} f_{n-1}(y)  \nonumber  \\
&& \hspace{-40mm}
 + \, a_{4}(\pm \beta) \Big\{ \left( \lambda_{\pm \beta} + \ln x \right) \left( \frac{1}{x} f_{0}(y) + \frac{4}{x^{2}} f_{1}(y) \right)  \nonumber  \\
&& \hspace{-24mm}
 - \frac{1}{x} f_{0,\ln}(y) - \frac{4}{x^{2}} f_{1,\ln}(y) \nonumber  \\
&& \hspace{-24mm}
 - \frac{2}{x} f_{0}(y) - \frac{4}{x^{2}} f_{1}(y)  \Big\}  \bigg] \, ,
\label{eq2-2-10}
\end{eqnarray}
\begin{eqnarray}
H_{1,\pm}(y) =  \alpha \bigg[ \sum_{n=0}^{3} \, b_{n}(\pm \beta) \, \frac{n^{2}}{x^{n}} f_{n-1}(y)  \nonumber  \\
&& \hspace{-40mm}
 + \, b_{4}(\pm \beta) \Big\{ \left( \lambda_{\pm \beta} + \ln x \right) \frac{9}{x^{3}} f_{2}(y)  \nonumber  \\
&& \hspace{-24mm}
- \frac{6}{x^{3}} f_{2}(y) - \frac{9}{x^{3}} f_{2,\ln}(y) \Big\}  \nonumber \\
&& \hspace{-40mm}
 + \, b_{5}(\pm \beta) \Big\{ \left( \lambda_{\pm \beta} + \ln x \right) \left( \frac{1}{x} f_{0}(y) + \frac{4}{x^{2}} f_{1}(y) \right)  \nonumber  \\
&& \hspace{-24mm}
 - \frac{1}{x} f_{0,\ln}(y) - \frac{4}{x^{2}} f_{1,\ln}(y) \nonumber  \\
&& \hspace{-24mm}
 - \frac{2}{x} f_{0}(y) - \frac{4}{x^{2}} f_{1}(y)  \Big\}  \bigg] \, ,
\label{eq2-2-11}
\end{eqnarray}
\begin{eqnarray}
H_{2,\pm}(y) =  \alpha \bigg[ \sum_{n=-1}^{4} \, c_{n}(\pm \beta) \, \frac{n^{2}}{x^{n}} f_{n-1}(y)  \nonumber  \\
&& \hspace{-40mm}
 + \, c_{5}(\pm \beta) \Big\{ \left( \lambda_{\pm \beta} + \ln x \right) \frac{9}{x^{3}} f_{2}(y)  \nonumber  \\
&& \hspace{-24mm}
- \frac{6}{x^{3}} f_{2}(y) - \frac{9}{x^{3}} f_{2,\ln}(y) \Big\}  \nonumber \\
&& \hspace{-40mm}
 + \, c_{6}(\pm \beta) \Big\{ \left( \lambda_{\pm \beta} + \ln x \right) \left( \frac{1}{x} f_{0}(y) + \frac{4}{x^{2}} f_{1}(y) \right)  \nonumber  \\
&& \hspace{-24mm}
 - \frac{1}{x} f_{0,\ln}(y) - \frac{4}{x^{2}} f_{1,\ln}(y) \nonumber  \\
&& \hspace{-24mm}
 - \frac{2}{x} f_{0}(y) - \frac{4}{x^{2}} f_{1}(y)  \Big\}  \bigg] \, ,
\label{eq2-2-12}
\end{eqnarray}
\begin{eqnarray}
H_{\ell,\rm{def}}(x) = - \frac{3}{4\beta^{2}} \left[ (1-\beta^{2}) \, n(x) - \frac{1}{2} (-1)^{\ell} (1+\beta)^{3} \, n(\xi x) \right.  \nonumber  \\
&& \hspace{-60mm}
\left. - \frac{1}{2} (-1)^{\ell} (1-\beta)^{3} \, n(x/\xi) \right]  \, .
\label{eq2-2-13}
\end{eqnarray}
Note that equation (\ref{eq2-2-13}) is the non-vanishing definite integral term in the integration by parts of equation (\ref{eq2-1-4}), where we used equations (\ref{eqA-29})--(\ref{eqA-33}) in the derivation.

  Thus, analytic expressions for $F_{0}(\beta,x)$, $G_{\ell}(\beta,x)$ and $H_{\ell}(\beta,x)$ are obtained provided that the functions $f_{n}(y)$ and $f_{n,\ln}(y)$ are known explicitly.

\subsection{Analytic structure of $f_{n}(y)$ and $f_{n,\ln}(y)$}

  In order to derive analytic forms for $f_{n}(y)$ and $f_{n,\ln}(y)$, we use the same method as \cite{zdzi13}, introducing
\begin{eqnarray}
\frac{1}{{\rm{e}}^{u} -1} = \left\{
\begin{array}{ll}
\displaystyle{\frac{1}{u} - \frac{1}{2} + \sum_{k=1}^{\infty} \frac{B_{2k} \, u^{2k-1}}{(2k)!}} &\, \, \,  {\rm for} \, \, \, 0 < u < 2\pi \\
\displaystyle{\sum_{k=1}^{\infty} {\rm{e}}^{-k u}}  & \, \, \,  {\rm for} \, \, \, u > 0    \\
\end{array}
\right.  \, ,
\label{eq2-3-1}
\end{eqnarray}
where $B_{2k}$ is the Bernoulli number.  Then, one can rewrite equation (\ref{eq2-2-3}) as follows:
\begin{eqnarray}
f_{n}(y) = \left\{
\begin{array}{ll}
g_{n0} \, - \, g_{n1}(y)  & \, \, \,  {\rm for} \, \, \, y < q \\
 \\
g_{n2}(y)  & \, \, \,  {\rm for} \, \, \, y > q    \\
\end{array}
\right.  \, ,
\label{eq2-3-2}
\end{eqnarray}
where $q$ is a number of the order of 2.  In \cite{zdzi13}, a value $q=2.257$ was chosen to minimize the maximum error in their calculation.  We have numerically checked the independence of $f_{n}(y)$ and $f_{n,\ln}(y)$ on the values of $q$, and we also use $q=2.257$ in this paper.

  For $n=-2$, one has
\begin{eqnarray}
g_{-20} = g_{-21}(q) + \int_{q}^{\infty} {\rm{d}}u \frac{u^{-2}}{{\rm{e}}^{u}-1} = 0.20065317162532  \, ,
\label{eq2-3-3}  \\
&&\hspace{-84mm}
g_{-21}(y) = -\frac{1}{2y^{2}} + \frac{1}{2y} + \frac{1}{2} B_{2} \ln y + \sum_{k=2}^{\infty} \frac{B_{2k} \, y^{2k-2}}{(2k-2)(2k)!}  \, ,
\label{eq2-3-4}  \\
&&\hspace{-84mm}
g_{-22}(y) = \frac{1}{y} \frac{1}{{\rm{e}}^{y}-1} - \sum_{k=1}^{\infty} k \, E_{1}(ky)  \, ,
\label{eq2-3-5}
\end{eqnarray}
where $E_{1}(y)=\int_{y}^{\infty}{\rm{d}}u \, {\rm{e}}^{-u}/u$ is the exponential integral.  Similarly, one obtains for $n=-1$,
\begin{eqnarray}
g_{-10} = g_{-11}(q) + \int_{q}^{\infty} {\rm{d}}u \frac{u^{-1}}{{\rm{e}}^{u}-1} = -0.63033070075391  \, ,
\label{eq2-3-6}  \\
&&\hspace{-86mm}
g_{-11}(y) = -\frac{1}{y} - \frac{1}{2} \ln y + \sum_{k=1}^{\infty} \frac{B_{2k} \, y^{2k-1}}{(2k-1)(2k)!}  \, ,
\label{eq2-3-7}  \\
&&\hspace{-86mm}
g_{-12}(y) = \sum_{k=1}^{\infty} \, E_{1}(ky)  \, .
\label{eq2-3-8}
\end{eqnarray}
Equation (\ref{eq2-2-3}) is integrated analytically for $n=0$, which is
\begin{eqnarray}
f_{0}(y) = -\ln(1-{\rm{e}}^{-y})  \, .
\label{eq2-3-9}
\end{eqnarray}
For $n \geq 1$, one has the following forms:
\begin{eqnarray}
g_{n0} = \Gamma(n+1) \, \zeta(n+1)  \, ,
\label{eq2-3-10}  \\
&&\hspace{-43mm}
g_{n1}(y) = \frac{1}{n} y^{n} - \frac{1}{2(n+1)} y^{n+1} + \sum_{k=1}^{\infty} \frac{B_{2k} \, y^{2k+n}}{(2k+n)(2k)!}  \, ,
\label{eq2-3-11}  \\
&&\hspace{-43mm}
g_{n2}(y) = - y^{n} \ln (1-{\rm{e}}^{-y}) + \sum_{m=1}^{n} \frac{n! \, y^{n-m}}{(n-m)!} \, L_{m+1}({\rm{e}}^{-y})  \, ,
\label{eq2-3-12}
\end{eqnarray}
where $\Gamma(n)$ and $\zeta(n)$ are the Gamma and Riemann Zeta functions, respectively, and $L_{m}(y) = \sum_{k=1}^{\infty}y^{k}/k^{m}$.

  Similarly, one can rewrite equation (\ref{eq2-2-4}) as follows:
\begin{eqnarray}
f_{n,\ln}(y) = \left\{
\begin{array}{ll}
h_{n0} \, - \, h_{n1}(y)  & \, \, \,  {\rm for} \, \, \, y < q \\
 \\
h_{n2}(y)  & \, \, \,  {\rm for} \, \, \, y > q    \\
\end{array}
\right.  \, .
\label{eq2-3-13}
\end{eqnarray}
For $n=0$, one has
\begin{eqnarray}
h_{00} = h_{01}(q) + \int_{q}^{\infty} {\rm{d}}u \frac{\ln u}{{\rm{e}}^{u}-1} = 0.72869391700393  \, ,
\label{eq2-3-14}  \\
&&\hspace{-80mm}
h_{01}(y) = \frac{1}{2}(\ln y)^{2} - \frac{1}{2} y \left(\ln y-1 \right)  \nonumber  \\
&&\hspace{-69mm}
 + \ln y \sum_{k=1}^{\infty} \frac{B_{2k} \, y^{2k}}{2k(2k)!} - \sum_{k=1}^{\infty} \frac{B_{2k} \, y^{2k}}{(2k)^{2}(2k)!}  \, ,
\label{eq2-3-15}  \\
&&\hspace{-80mm}
h_{02}(y) = - \ln y \, \ln(1-{\rm{e}}^{-y}) + \sum_{k=1}^{\infty} \frac{1}{k} \, E_{1}(ky)  \, .
\label{eq2-3-16}
\end{eqnarray}
For $n \geq 1$, one obtains
\begin{eqnarray}
h_{n0} = \Gamma^{\prime}(n+1) \, \zeta(n+1) + \Gamma(n+1) \, \zeta^{\prime}(n+1) \, ,
\label{eq2-3-17}  \\
\nonumber  \\
&&\hspace{-73mm}
h_{n1}(y) = \frac{1}{n} y^{n} \left( \ln y - \frac{1}{n} \right) - \frac{1}{2(n+1)} y^{n+1} \left( \ln y - \frac{1}{n+1} \right) \nonumber  \\
&&\hspace{-62mm}
 + \ln y \sum_{k=1}^{\infty} \frac{B_{2k} \, y^{2k+n}}{(2k+n)(2k)!} - \sum_{k=1}^{\infty} \frac{B_{2k} \, y^{2k+n}}{(2k+n)^{2}(2k)!}  \, ,
\label{eq2-3-18}  \\
&&\hspace{-73mm}
h_{n2}(y) = - y^{n} \ln y \, \ln (1-{\rm{e}}^{-y}) + \, n! \sum_{k=1}^{\infty} \frac{1}{k^{n+1}} \, E_{1}(ky)   \nonumber  \\
&&\hspace{-62mm}
+ \, \ln y \sum_{m=1}^{n} \frac{n! \, y^{n-m}}{(n-m)!} \, L_{m+1}({\rm{e}}^{-y}) + \, \sum_{m=0}^{n-1} \frac{n!}{(n-m)!}  \nonumber  \\
&& \hspace{-57mm}
\times \sum_{\ell=1}^{n-m} \frac{(n-m-1)!}{(n-m-\ell)!} \, y^{n-m-\ell} \, L_{m+\ell+1}({\rm{e}}^{-y})  \, ,
\label{eq2-3-19}
\end{eqnarray}
where $\Gamma^{\prime}(n)$ and $\zeta^{\prime}(n)$ are the derivatives of the Gamma and Reimann Zeta functions, respectively.

  Finally, we have checked the numerical convergence of the functions $f_{n}(y)$ and $f_{n,\ln}(y)$ by increasing the maximum number of summations $k$ in equations (\ref{eq2-3-4})--(\ref{eq2-3-19}), which include the sum in $L_{m}(y)$ as well.  It has been found that they converge very quickly as $k$ increases.  We summarize the results, for example, for $f_{-2}(y)$ and $f_{0,\ln}(y)$ in Table 1.  In Table 1, we list the $k$-values which give the accuracy of required number of significant digits at given positions $y$.  At $y=q$, the calculations with $k=14$ give the results accurate 14 significant digits for these functions.  It is also seen from Table 1 that these functions converge extremely fast for $y \geq 10$ region, where $k=4$ gives the accuracy of the 14 significant digits.  It should be noted that $k=3$ was used in \cite{zdzi13}, whereas we use $k=14$ in this paper.  The present calculation requires more accurate values for $f_{n}(y)$ and $f_{n,\ln}(y)$ in order to cope with the cancellations of significant digits in the numerical calculation of the functions $F_{0}(\beta,x)$, $G_{\ell}(\beta,x)$ and $H_{\ell}(\beta,x)$ at $\beta \rightarrow 0$, which will be discussed in detail in the following sections.  In Section 5, we study numerically the analytic formulas of the integrated photon redistribution functions for their applicable $(\beta,x)$ region.

\begin{table}
\caption[]{Numerical convergence of $f_{-2}(y)$ and $f_{0,\ln}(y)$.}
\begin{tabular}{ccccccc} \hline \hline

     &  number of  &       &       &  positions  &         &        \\
     &  digits  &  $y$=1  &  $y$=$q$  &  $y$=5  &  $y$=10  &  $y$=15  \\ \hline

     &  4 digits  &  $k$=2  &  $k$=4   &  $k$=3  &  $k$=2  &  $k$=1  \\
     &  6 digits  &  $k$=3  &  $k$=7   &  $k$=4  &  $k$=2  &  $k$=2  \\
$f_{-2}(y)$  &  8 digits  &  $k$=5  &  $k$=9   &  $k$=4  &  $k$=3  &  $k$=2  \\
     & 10 digits  &  $k$=5  &  $k$=11  &  $k$=5  &  $k$=3  &  $k$=2  \\
     & 12 digits  &  $k$=7  &  $k$=13  &  $k$=6  &  $k$=3  &  $k$=2  \\
     & 14 digits  &  $k$=9  &  $k$=14  &  $k$=7  &  $k$=4  &  $k$=3  \\ \hline \hline

     &  number of  &       &       &  positions  &         &        \\
     &  digits  &  $y$=1  &  $y$=$q$  &  $y$=5  &  $y$=10  &  $y$=15  \\ \hline

     &  4 digits  &  $k$=2  &  $k$=4   &  $k$=1  &  $k$=1  &  $k$=1  \\
     &  6 digits  &  $k$=3  &  $k$=7   &  $k$=2  &  $k$=1  &  $k$=1  \\
$f_{0,\ln}(y)$  &  8 digits  &  $k$=5  &  $k$=9   &  $k$=3  &  $k$=2  &  $k$=1  \\
     & 10 digits  &  $k$=5  &  $k$=11  &  $k$=4  &  $k$=2  &  $k$=2  \\
     & 12 digits  &  $k$=7  &  $k$=13  &  $k$=5  &  $k$=3  &  $k$=2  \\
     & 14 digits  &  $k$=9  &  $k$=14  &  $k$=6  &  $k$=3  &  $k$=2  \\ \hline \hline

\end{tabular}
\end{table}

\section{Power series expansion approximation}

  Numerical calculations of the functions $F_{0}(\beta,x)$, $G_{\ell}(\beta,x)$ and $H_{\ell}(\beta,x)$ are straightforward with the analytic forms of equations (\ref{eq2-2-2}), (\ref{eq2-2-6})--(\ref{eq2-2-8}) and (\ref{eq2-2-10})--(\ref{eq2-2-13}).  As will be shown explicitly in Section 5, however, the numerical errors become non-negligible in the region $\beta \rightarrow$ 0 because of the cancellations among various terms.  In the calculation of the function $F_{0}(\beta,x)$, for example, each coefficient diverges as $a_{n}(\beta) \sim \mathcal{O}(1/\beta^{4})$ and the cancellation of significant digits occurs at $\beta < 5.0 \times 10^{-3}$ for $k=14$ case with the double precision calculation.  For the calculation with smaller values of $k$, the cancellation of significant digits occurs at larger $\beta$-values as anticipated.  In the calculations of $G_{1}(\beta,x)$ and $H_{1}(\beta,x)$, each coefficient diverges as $b_{n}(\beta) \sim \mathcal{O}(1/\beta^{6})$.  Similarly, in the calculations of $G_{2}(\beta,x)$ and $H_{2}(\beta,x)$, each coefficient diverges as $c_{n}(\beta) \sim \mathcal{O}(1/\beta^{8})$.  Therefore, the cancellations of significant digits occur at larger $\beta$-values for these functions.  In order to avoid the numerical instability at $\beta \rightarrow 0$, it is indispensable to derive formulas which are more suitable for the region $\beta \ll 1$.

  Expanding $\xi=(1+\beta)/(1-\beta)$ in terms of the power series of $\beta$, one obtains approximate expressions for the integrated photon redistribution functions which are valid for $\beta \ll 1$.  The calculation is lengthy but straightforward.  It turns out that all contributions from inverse powers in $\beta$ vanish completely.  They can be written in a unified form.

  We summarize the results up to $\mathcal{O}(\beta^{10})$ as follows:
\begin{eqnarray}
F_{0}(\beta,x) \approx \sum_{k=0}^{5} \frac{1}{(2k+1)!!} \, \beta^{2k} A_{k} \, n(x)  \, ,
\label{eq3-1-1}  \\
&& \hspace{-61mm}
G_{0}(\beta,x) \approx \sum_{k=0}^{5} \frac{1}{(2k+1)!!} \, \beta^{2k} A_{k} \, D \, n(x)  \, ,
\label{eq3-1-2}  \\
&& \hspace{-61mm}
H_{0}(\beta,x) \approx \sum_{k=0}^{5} \frac{1}{(2k+1)!!} \, \beta^{2k} A_{k} \, D^{2} \, n(x)  \, ,
\label{eq3-1-3}  \\
&& \hspace{-61mm}
A_{0} = 1  \, ,
\label{eq3-1-4}  \\
&& \hspace{-61mm}
A_{1} = \Delta  \, ,
\label{eq3-1-5}  \\
&& \hspace{-61mm}
A_{2} = \frac{11}{5} \Delta + \frac{7}{10} \Delta^{2}  \, ,
\label{eq3-1-6}  \\
&& \hspace{-61mm}
A_{3} = \frac{157}{15} \Delta + \frac{14}{3} \Delta^{2} + \frac{11}{30} \Delta^{3}  \, ,
\label{eq3-1-7}  \\
&& \hspace{-61mm}
A_{4} = \frac{2523}{35} \Delta + \frac{8021}{210} \Delta^{2} + \frac{211}{42} \Delta^{3} + \frac{16}{105} \Delta^{4}  \, ,
\label{eq3-1-8}  \\
&& \hspace{-61mm}
A_{5} = \frac{22539}{35} \Delta + \frac{40216}{105} \Delta^{2} + \frac{1375}{21} \Delta^{3} + \frac{374}{105} \Delta^{4}  \nonumber  \\
&& \hspace{-7mm}
+ \frac{11}{210} \Delta^{5} \, ,
\label{eq3-1-9}
\end{eqnarray}
where $\Delta = D(D+3)$ is the diffusion operator of the original Kompaneets equation \citep{komp56}.

  Similarly, one obtains
\begin{eqnarray}
G_{1}(\beta,x) \approx - \sum_{k=0}^{5} \frac{1}{(2k+1)!!} \, \beta^{2k} B_{k} \, D \, n(x)  \, ,
\label{eq3-1-10}  \\
&& \hspace{-68mm}
H_{1}(\beta,x) \approx - \sum_{k=0}^{5} \frac{1}{(2k+1)!!} \, \beta^{2k} B_{k} \, D^{2} \, n(x)  \, ,
\label{eq3-1-11}  \\
&& \hspace{-68mm}
B_{0} = 0  \, ,
\label{eq3-1-12}  \\
&& \hspace{-68mm}
B_{1} = \frac{2}{5} + \frac{2}{5} \Delta  \, ,
\label{eq3-1-13}  \\
&& \hspace{-68mm}
B_{2} = \frac{6}{5} + \frac{3}{5} \Delta + \frac{2}{5} \Delta^{2}  \, ,
\label{eq3-1-14}  \\
&& \hspace{-68mm}
B_{3} = \frac{138}{35} + \frac{26}{7} \Delta + \frac{78}{35} \Delta^{2} + \frac{17}{70} \Delta^{3} \, ,
\label{eq3-1-15}  \\
&& \hspace{-68mm}
B_{4} = \frac{678}{35} + \frac{427}{15} \Delta + \frac{1802}{105} \Delta^{2} + \frac{91}{30} \Delta^{3} + \frac{23}{210} \Delta^{4}  \, ,
\label{eq3-1-16}  \\
&& \hspace{-68mm}
B_{5} = \frac{4574}{35} + \frac{83366}{315} \Delta + \frac{53008}{315} \Delta^{2} + \frac{3917}{105} \Delta^{3}  \nonumber  \\
&& \hspace{-19mm}
  + \frac{109}{45} \Delta^{4} + \frac{5}{126} \Delta^{5}  \, ,
\label{eq3-1-17}
\end{eqnarray}
and
\begin{eqnarray}
G_{2}(\beta,x) \approx \sum_{k=0}^{5} \frac{1}{(2k+1)!!} \, \beta^{2k} C_{k} \, D \, n(x)  \, ,
\label{eq3-1-18}  \\
&& \hspace{-65mm}
H_{2}(\beta,x) \approx \sum_{k=0}^{5} \frac{1}{(2k+1)!!} \, \beta^{2k} C_{k} \, D^{2} \, n(x)  \, ,
\label{eq3-1-19}  \\
&& \hspace{-65mm}
C_{0} = \frac{1}{10}  \, ,
\label{eq3-1-20}  \\
&& \hspace{-65mm}
C_{1} = - \frac{3}{5} + \frac{1}{10} \Delta  \, ,
\label{eq3-1-21}  \\
&& \hspace{-65mm}
C_{2} = \frac{39}{35} - \frac{1}{2} \Delta + \frac{1}{7} \Delta^{2}  \, ,
\label{eq3-1-22}  \\
&& \hspace{-65mm}
C_{3} = \frac{81}{35} - \frac{5}{42} \Delta + \frac{22}{105} \Delta^{2} + \frac{23}{210} \Delta^{3}  \, ,
\label{eq3-1-23}  \\
&& \hspace{-65mm}
C_{4} = \frac{201}{35} + \frac{181}{42} \Delta + \frac{103}{105} \Delta^{2} + \frac{209}{210} \Delta^{3} + \frac{2}{35} \Delta^{4}  \, ,
\label{eq3-1-24}  \\
&& \hspace{-65mm}
C_{5} = \frac{69}{5} + \frac{287}{6} \Delta + \frac{6028}{525} \Delta^{2} + \frac{5191}{525} \Delta^{3} + \frac{38}{35} \Delta^{4}  \nonumber  \\
&& \hspace{-8mm}
 + \frac{4}{175} \Delta^{5}   \, .
\label{eq3-1-25}
\end{eqnarray}
We analyse numerically the applicable $(\beta,x)$ region of these formulas in Section 5.

  Finally, inserting equations (\ref{eq3-1-1})--(\ref{eq3-1-3}), (\ref{eq3-1-10})--(\ref{eq3-1-11}) and (\ref{eq3-1-18})--(\ref{eq3-1-19}) into equation (\ref{eq2-1-1}), one can reproduce the result of the Fokker$-$Planck expansion approximation obtained in Appendix B1 of \cite{noza13} for the thermal electron distribution.

\section{Calculation in the CMB and general OBS frames}

  The Boltzmann equations in the CMB and OBS frames were given by equations (38) and (43) of \cite{noza13}, respectively, and the following photon redistribution functions appeared in these frames:
\begin{equation}
P_{m}(s,\beta) = P_{0}(s,\beta)  \, ,
\label{eq4-1-1}
\end{equation}
\begin{equation}
P_{d}(s,\beta) = P_{0}(s,\beta) - P_{1}(s,\beta)  \, ,
\label{eq4-1-2}
\end{equation}
\begin{equation}
P_{q}(s,\beta) = P_{0}(s,\beta) - 2 \, P_{1}(s,\beta) + P_{2}(s,\beta)  \, .
\label{eq4-1-3}
\end{equation}

  The integrated photon redistribution functions for equations (\ref{eq4-1-1})--(\ref{eq4-1-3}) are given by
\begin{eqnarray}
F_{m}(\beta,x) = \int_{-\lambda_{\beta}}^{+\lambda_{\beta}} {\rm{d}}s \, P_{m}(s,\beta) \, n({\rm{e}}^{s}x) = F_{0}(\beta,x)  \, ,
\label{eq4-1-4}  \\
&& \hspace{-76mm}
G_{d}(\beta,x) = \int_{-\lambda_{\beta}}^{+\lambda_{\beta}} {\rm{d}}s \, P_{d}(s,\beta) \, D \, n({\rm{e}}^{s}x)  \nonumber  \\
&& \hspace{-63mm}
= G_{0}(\beta,x) - G_{1}(\beta,x)  \, ,
\label{eq4-1-5}  \\
&& \hspace{-76mm}
G_{q}(\beta,x) = \int_{-\lambda_{\beta}}^{+\lambda_{\beta}} {\rm{d}}s \, P_{q}(s,\beta) \, D \, n({\rm{e}}^{s}x)  \nonumber  \\
&& \hspace{-63mm}
= G_{0}(\beta,x) - 2 \, G_{1}(\beta,x) + G_{2}(\beta,x)  \, ,
\label{eq4-1-6}  \\
&& \hspace{-76mm}
H_{d}(\beta,x) = \int_{-\lambda_{\beta}}^{+\lambda_{\beta}} {\rm{d}}s \, P_{d}(s,\beta) \, D^{2} \, n({\rm{e}}^{s}x)  \nonumber  \\
&& \hspace{-63mm}
= H_{0}(\beta,x) - H_{1}(\beta,x)  \, ,
\label{eq4-1-7}  \\
&& \hspace{-76mm}
H_{q}(\beta,x) = \int_{-\lambda_{\beta}}^{+\lambda_{\beta}} {\rm{d}}s \, P_{q}(s,\beta) \, D^{2} \, n({\rm{e}}^{s}x)  \nonumber  \\
&& \hspace{-63mm}
= H_{0}(\beta,x) - 2 \, H_{1}(\beta,x) + H_{2}(\beta,x)  \, .
\label{eq4-1-8}
\end{eqnarray}
Thus, the Boltzmann equations in the CMB and OBS frames are calculated with equations (\ref{eq4-1-4})--(\ref{eq4-1-8}).

\section{Numerical analyses on the integrated photon redistribution functions}

  In this section, we study numerically the analytic formulas and the power series expansion approximation of the integrated photon redistribution functions.  Especially, we explore the applicable $(\beta,x)$ region of these formulas.

  Let us start with the $\mathcal{O}(\beta_{c}^{0})$ function in equation (\ref{eq2-1-1}).  In Figs.~1a and 1b, we plot $F_{0}(\beta,x)$ as a function of $\beta$ at $x=2.3$ and $x=7.2$, respectively.  Note that these points correspond to the minimum and maximum positions of the thermal SZ effect spectrum.  In Figs.~1c and 1d, we plot $F_{0}(\beta,x)$ in higher frequency region, for example, at $x=10$ and $x=15$.  In these figures, the solid curve corresponds to the analytic formula and the dotted curve corresponds to the power series expansion approximation.  At small $x$ region, two calculations agree extremely well for a wide range of $\beta$ values.  As $x$ increases, however, the applicable region of the power series expansion approximation is restricted to smaller $\beta$ values as anticipated.  It can be concluded from Fig.~1 that the analytic formula is applicable to $0.04 \leq \beta$ for $x \leq 15$ region, and the power series expansion approximation is applicable to $\beta \leq 0.5$ for $x \leq 15$ region.  In the overlapping region $0.04 \leq \beta \leq 0.5$, two calculations give an excellent agreement of more than four significant digits.

\begin{figure}
\begin{center}
\includegraphics[angle=0,width=0.44\textwidth]{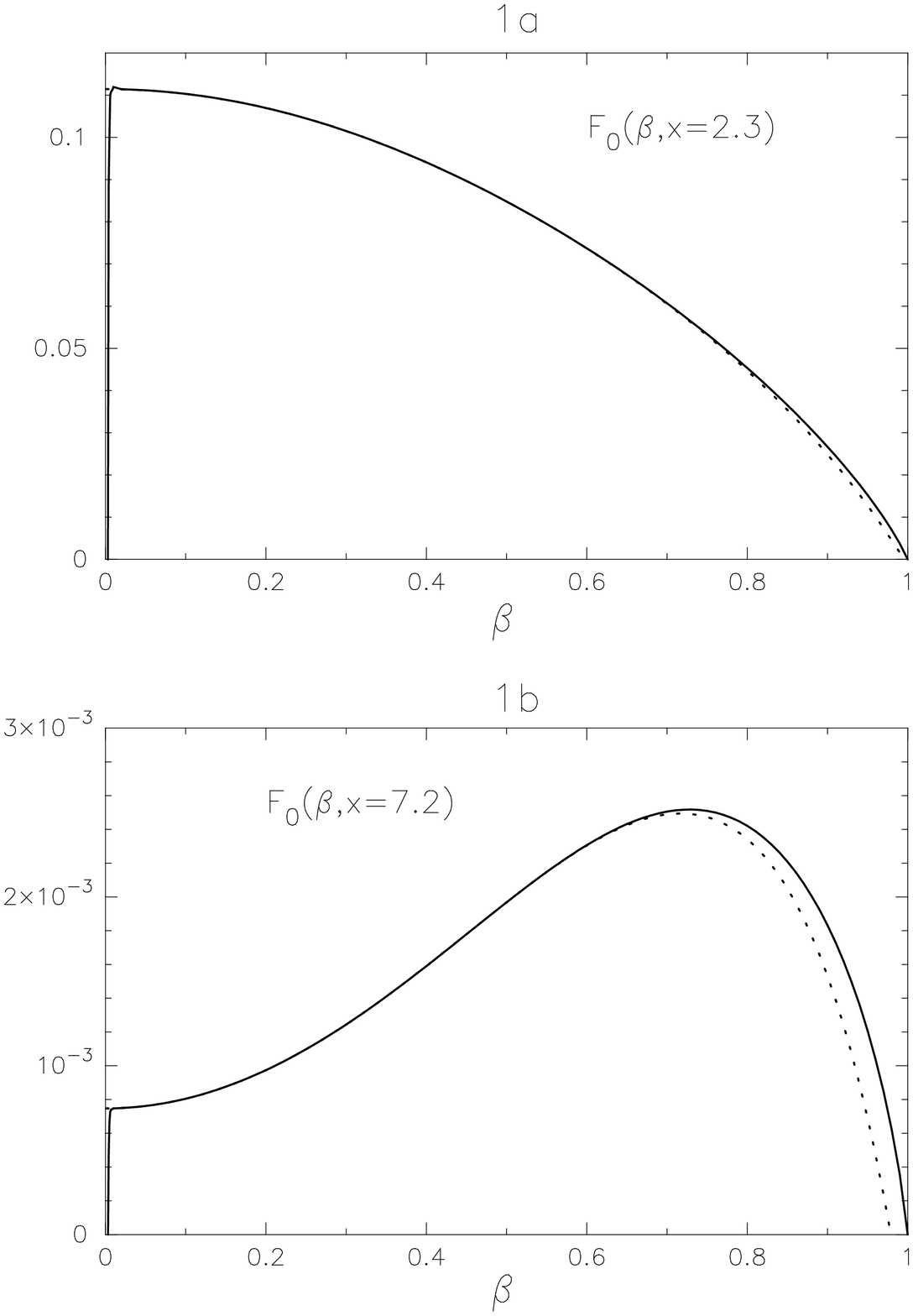}
\includegraphics[angle=0,width=0.44\textwidth]{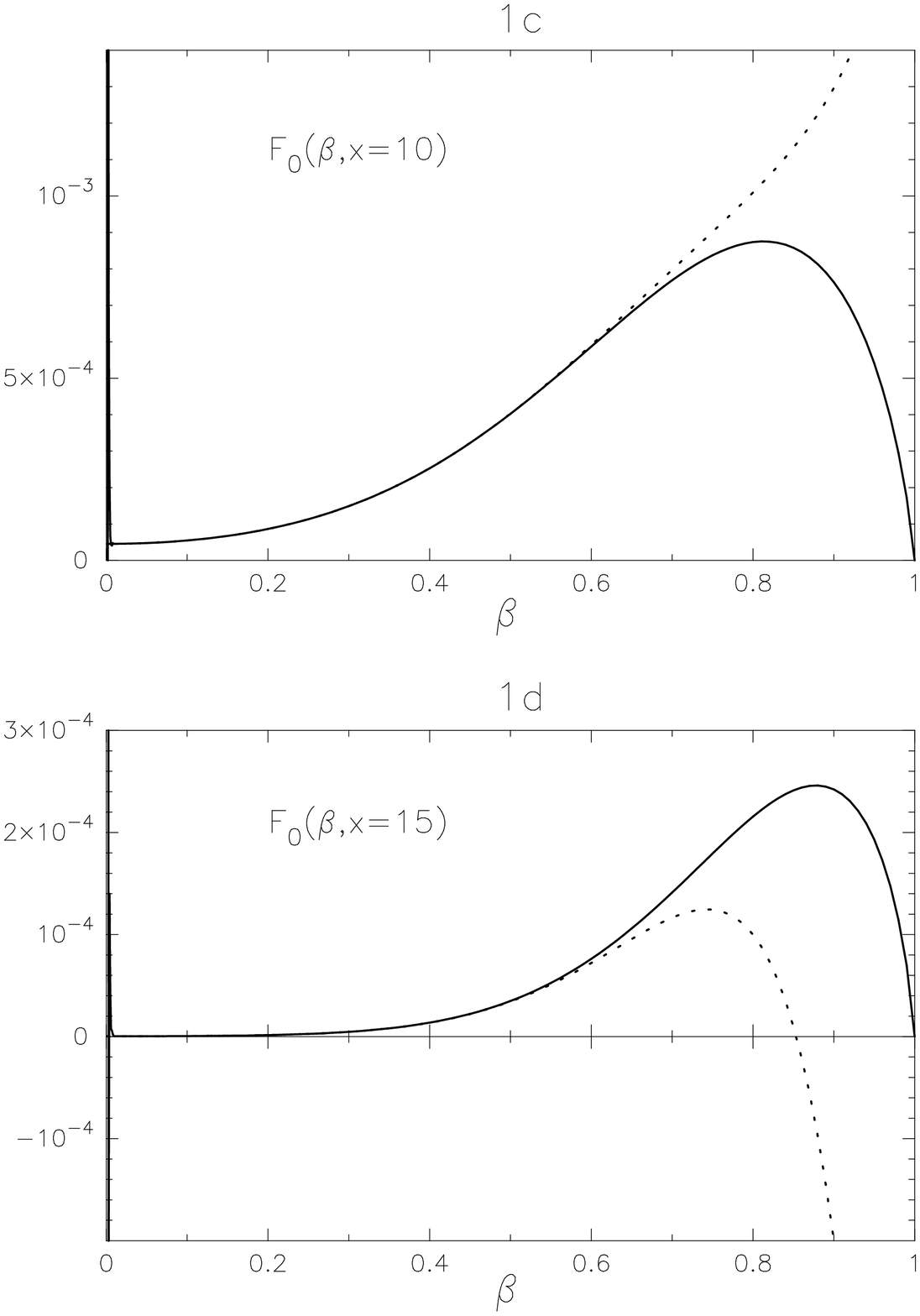}
\end{center}
\caption{Plotting of $F_{0}(\beta,x)$ as a function of $\beta$ at $x=2.3$, $x=7.2$, $x=10$ and $x=15$.  The solid curve corresponds to the analytic formula.  The dotted curve corresponds to the power series expansion approximation.}
\end{figure}

  We now plot the $\mathcal{O}(\beta_{c})$ functions.  In Figs.~2a and 2b, we plot $G_{0}(\beta,x)$ as a function of $\beta$ at $x=4$ and $x=15$, respectively.  Similarly, we plot $G_{1}(\beta,x)$ in Figs.~2c and 2d at $x=4$ and $x=15$.  Note that $x=4$ corresponds to the maximum position of the kinematical SZ effect spectrum.  It can be concluded from Fig.~2 that the analytic formula is applicable to $0.11 \leq \beta$ for $x \leq 15$ region, and the power series expansion approximation is applicable to $\beta \leq 0.26$ for $x \leq 15$ region.  In the overlapping region $0.11 \leq \beta \leq 0.26$, two calculations give an excellent agreement of more than four significant digits.

\begin{figure}
\begin{center}
\includegraphics[angle=0,width=0.44\textwidth]{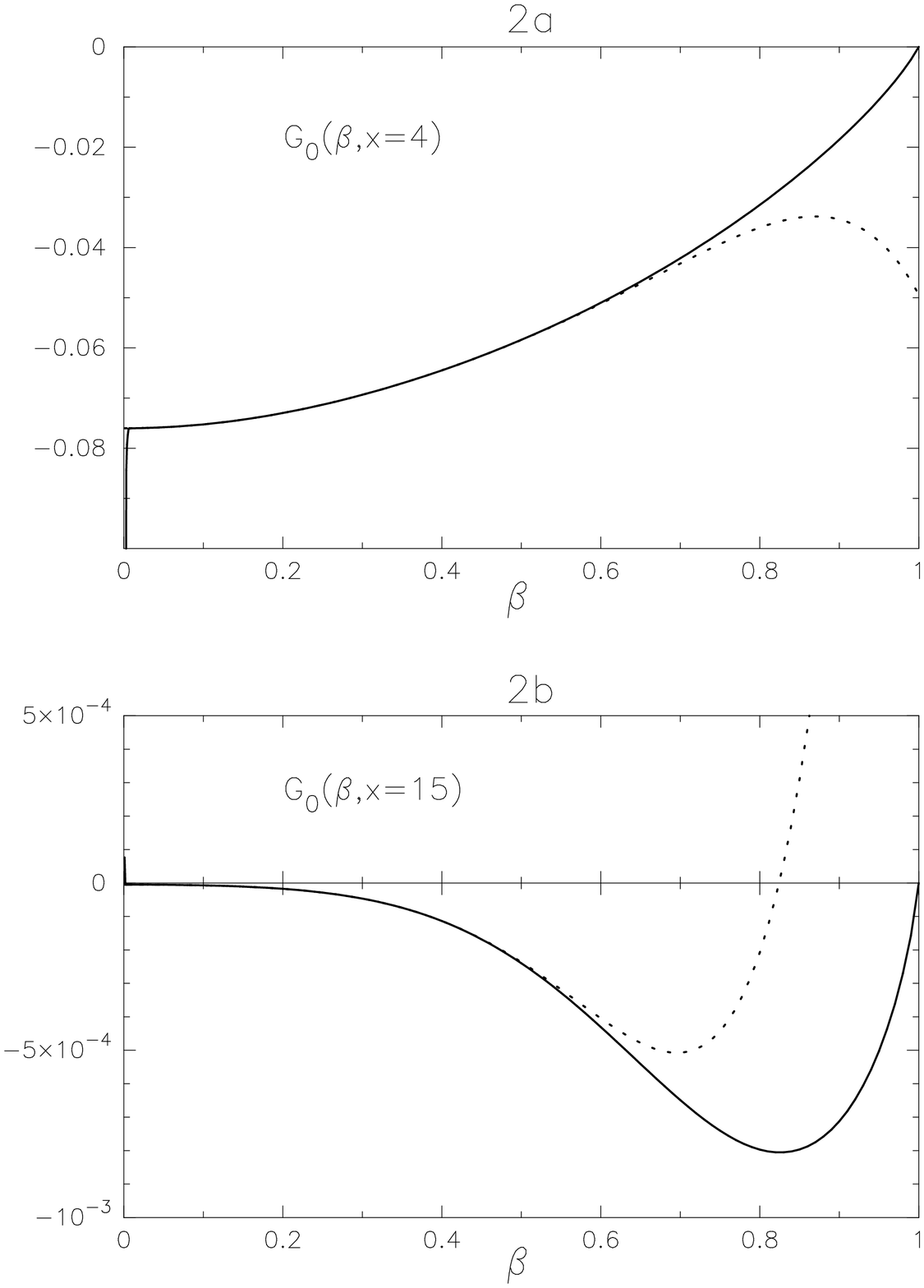}
\includegraphics[angle=0,width=0.44\textwidth]{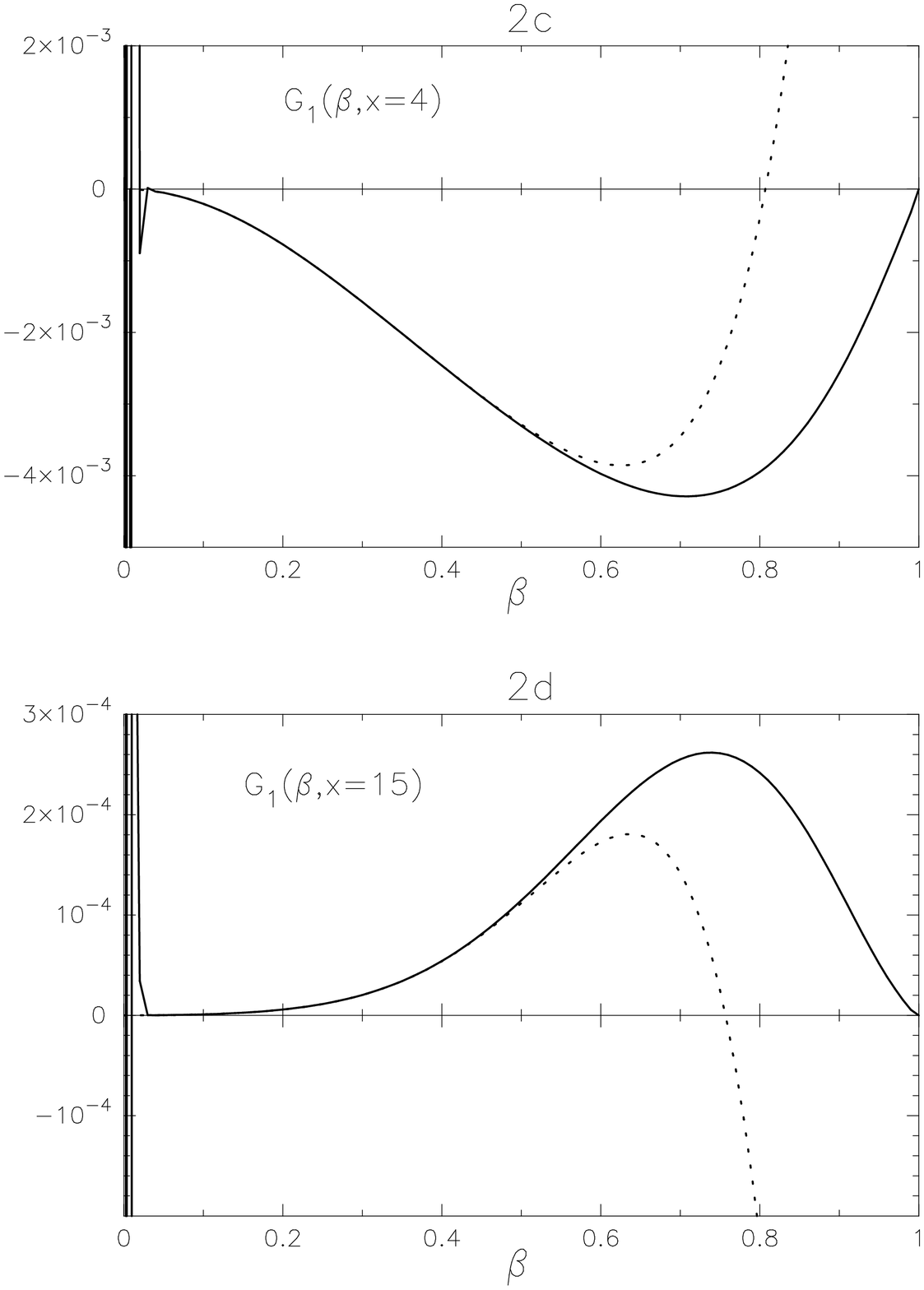}
\end{center}
\caption{Plotting of $G_{0}(\beta,x)$ and $G_{1}(\beta,x)$ as a function of $\beta$ at $x=4$ and $x=15$.  The solid curve corresponds to the analytic formula.  The dotted curve corresponds to the power series expansion approximation.}
\end{figure}

  Finally, we plot the $\mathcal{O}(\beta_{c}^{2})$ functions.  In Figs.~3a, 3b, 3c and 3d, we plot $H_{0}(\beta,x)$, $H_{1}(\beta,x)$, $H_{2}(\beta,x)$ and $G_{2}(\beta,x)$ as a function of $\beta$ at $x=15$.  It can be concluded from Fig.~3 that the analytic formula is applicable to $0.16 \leq \beta$ for $x \leq 15$ region, and the power series expansion approximation is applicable to $\beta \leq 0.20$ for $x \leq 15$ region.  In the overlapping region $0.16 \leq \beta \leq 0.20$, two calculations give an excellent agreement of more than four significant digits.

\begin{figure}
\begin{center}
\includegraphics[angle=0,width=0.44\textwidth]{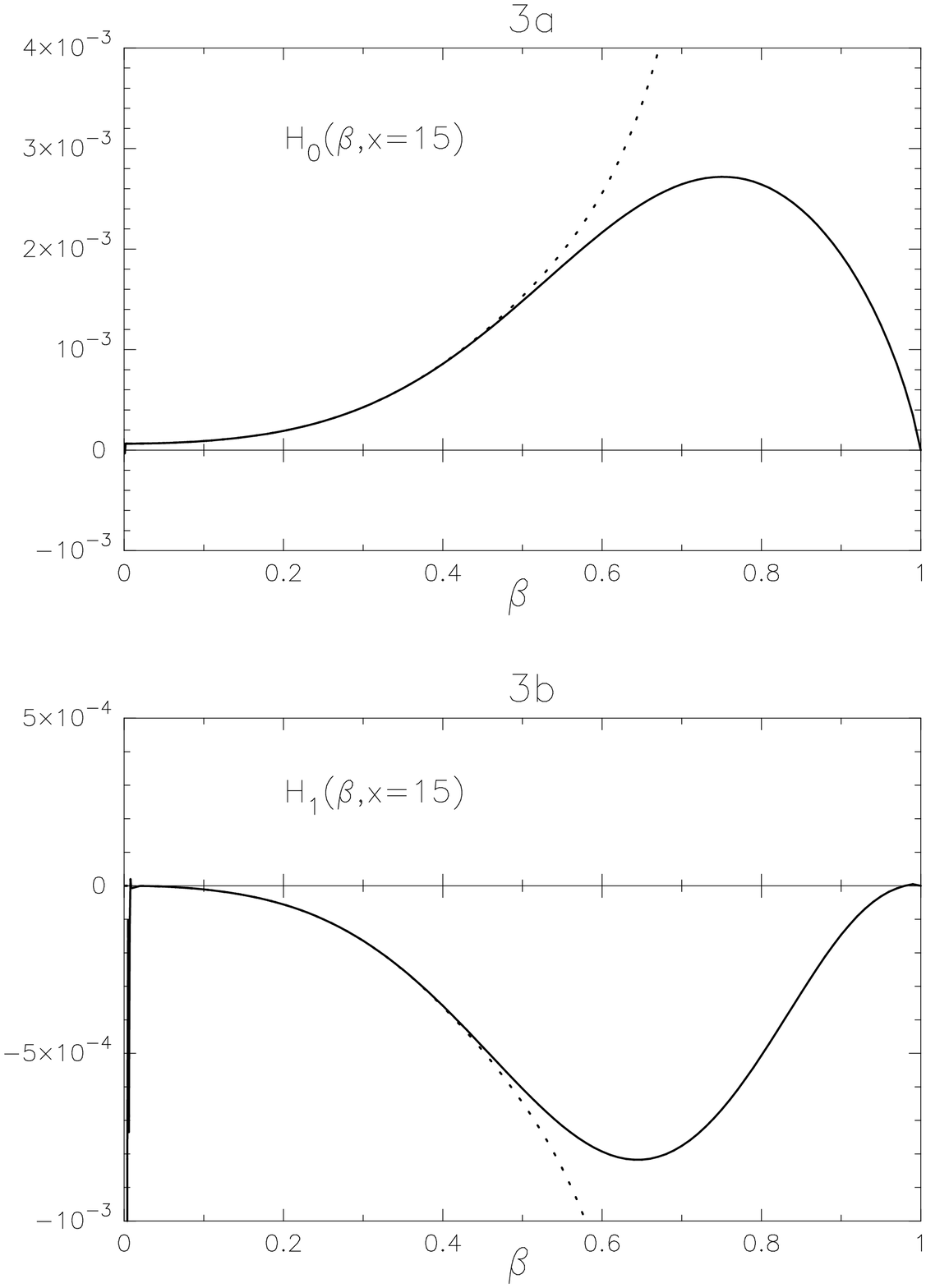}
\includegraphics[angle=0,width=0.44\textwidth]{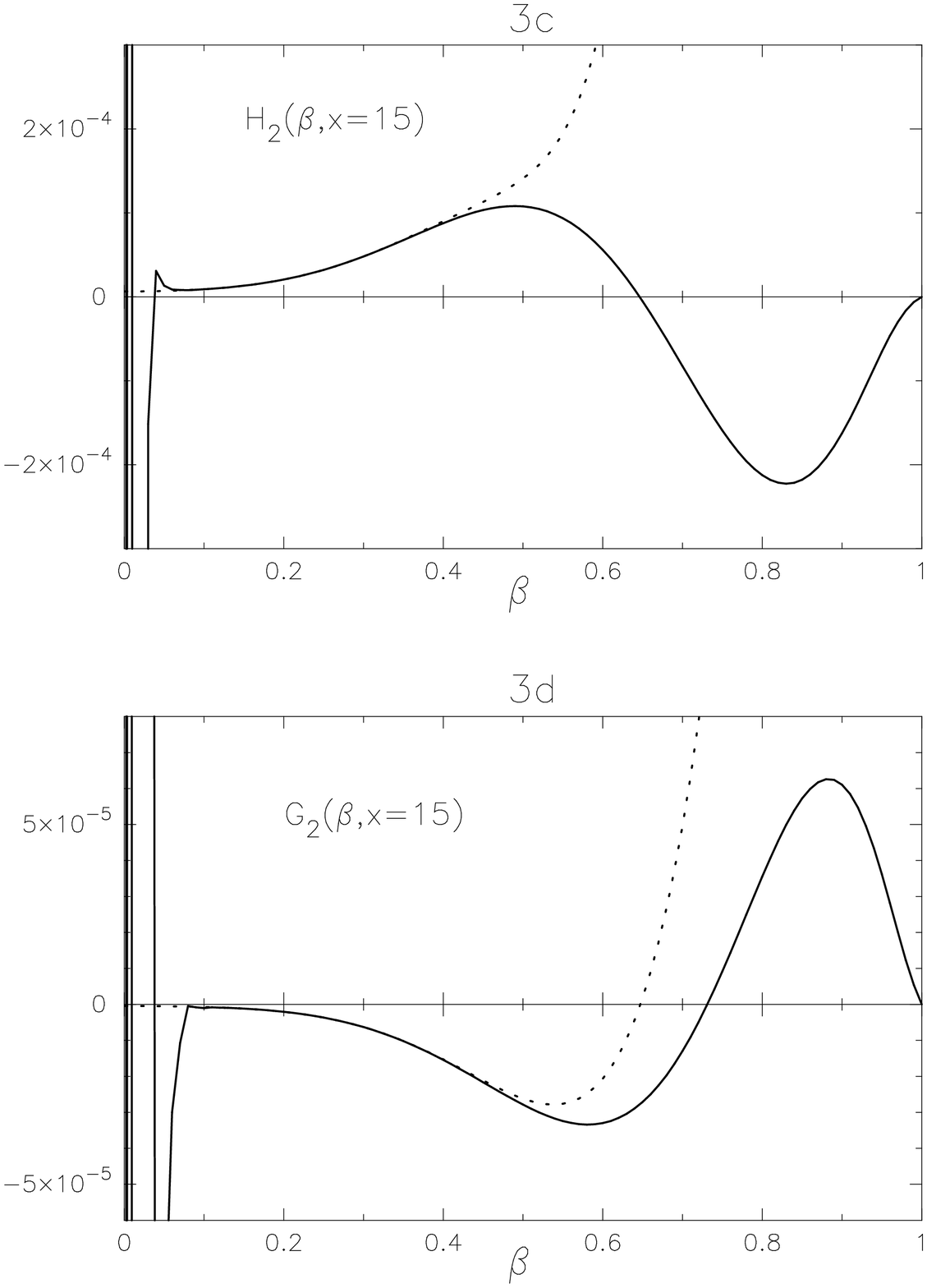}
\end{center}
\caption{Plotting of $H_{0}(\beta,x)$, $H_{1}(\beta,x)$, $H_{2}(\beta,x)$ and $G_{2}(\beta,x)$ as a function of $\beta$ at $x=15$.  The solid curve corresponds to the analytic formula.  The dotted curve corresponds to the power series expansion approximation.}
\end{figure}

  Thus, the analytic formulas for $F_{0}(\beta,x)$, $G_{\ell}(\beta,x)$ and $H_{\ell}(\beta,x)$ combined with the power series expansion approximation formulas provide us a simple and accurate numerical tool for the $0 \leq \beta \leq 1$, $ 0 \leq x \leq 15$ region to analyse observation data.  We summarize the results in Table 2.  In Table 2, $\beta_{*}$ denotes the boundary $\beta$-value of the two regions, where the power series approximation formula can be used for region I ($0 \leq \beta \leq \beta_{*}$, $0 \leq x \leq 15$), and the analytic formula can be used for region II ($\beta_{*} \leq \beta \leq 1$, $0 \leq x \leq 15$).  It should be noted that these formulas at $\beta=\beta_{*}$ have accuracy of more than four significant digits.  The Fortran numerical program is also available from one of authors (S.N.) upon request.

  Before closing this section, one should comment on the behavior of the analytic functions of $F_{0}(\beta,x)$, $G_{\ell}(\beta,x)$ and $H_{\ell}(\beta,x)$ at $\beta=1$.  As shown explicitly in equations (\ref{eqA-2})--(\ref{eqA-27}), the photon redistribution functions $P_{\ell}(s, \beta)=0$ at $\beta=1$ for any values of $s$, because $\alpha =0$, the coefficients $a_{n}(\beta)$, $b_{n}(\beta)$ and $c_{n}(\beta)$ are zero or finite at $\beta=1$.  Therefore, the integrated photon redistribution functions $F_{0}(\beta,x)$, $G_{\ell}(\beta,x)$ and $H_{\ell}(\beta,x)$ at $\beta=1$ are also zero for any values of $x$ by the definitions of equations (\ref{eq2-1-2})--(\ref{eq2-1-4}).  These features are clearly seen in Figs.~1--3.  Furthermore, we have examined the behavior of the analytic formula of $F_{0}(\beta,x)$ for $\beta \rightarrow 1$.  In equation (\ref{eq2-2-1}), $F_{0,+}(x) \sim \mathcal{O}(1-\beta)$, $F_{0,-}(x) \sim \mathcal{O}(1-\beta)$, $F_{0,+}(\xi x) \sim \mathcal{O}(\exp(-\xi x)/(1-\beta)^{m})$ and $F_{0,-}(x/\xi) \sim \mathcal{O}((1-\beta)^{2} \ln^{2}(1-\beta))$ for $\beta \rightarrow 1$.  Therefore,  $F_{0}(\beta,x) \sim \mathcal{O}(1-\beta)$ for $\beta \rightarrow 1$.  Behavior of $G_{\ell}(\beta,x)$ and $H_{\ell}(\beta,x)$ for $\beta \rightarrow 1$ is also obtained in a similar manner.  One obtains $G_{\ell}(\beta,x) \sim \mathcal{O}(1-\beta)$ and $H_{\ell}(\beta,x) \sim \mathcal{O}(1-\beta)$.

\begin{table}
\caption[]{Numerical recipes for $F_{0}(\beta,x)$, $G_{\ell}(\beta,x)$ and $H_{\ell}(\beta,x)$.}
\begin{tabular}{cccc} \hline \hline

     &              &     region I                      &  region II  \\
     &  $\beta_{*}$ &  \, $0 \leq \beta \leq \beta_{*}$ &  $\beta_{*} \leq \beta \leq 1$ \\   

     &              &  \, $0 \leq x \leq 15$   &  \, $0 \leq x \leq 15$ \\ \hline \hline

 $F_{0}(\beta,x)$  &  0.18  &  eq.~(\ref{eq3-1-1})  &  \, eqs.~(\ref{eq2-2-1}), (\ref{eq2-2-2})  \\

 $G_{0}(\beta,x)$  &  0.18  &  eq.~(\ref{eq3-1-2})  &  \, eqs.~(\ref{eq2-2-5}), (\ref{eq2-2-6})  \\

 $H_{0}(\beta,x)$  &  0.18  &  eq.~(\ref{eq3-1-3})  &  \, eqs.~(\ref{eq2-2-9}), (\ref{eq2-2-10}), (\ref{eq2-2-13}) \\ \hline

 $G_{1}(\beta,x)$  &  0.18  &  eq.~(\ref{eq3-1-10})  &  \, eqs.~(\ref{eq2-2-5}), (\ref{eq2-2-7})  \\

 $H_{1}(\beta,x)$  &  0.18  &  eq.~(\ref{eq3-1-11})  &  \, eqs.~(\ref{eq2-2-9}), (\ref{eq2-2-11}), (\ref{eq2-2-13})  \\ \hline

 $G_{2}(\beta,x)$  &  0.18  &  eq.~(\ref{eq3-1-18})  &  \, eqs.~(\ref{eq2-2-5}), (\ref{eq2-2-8})  \\

 $H_{2}(\beta,x)$  &  0.18  &  eq.~(\ref{eq3-1-19})  &  \, eqs.~(\ref{eq2-2-9}), (\ref{eq2-2-12}), (\ref{eq2-2-13})  \\ \hline \hline

\end{tabular}
\end{table}

\section{Conclusions}

We studied the SZ effect for the CG in the Thomson approximation.  In Section 2, we investigated the Boltzmann equation for the photon distribution function in the CG frame.  We introduced the integrated photon redistribution functions $F_{0}(\beta,x)$, $G_{\ell}(\beta,x)$ and $H_{\ell}(\beta,x)$, and expressed the Boltzmann equation in terms of these functions, where $x=\omega/k_{B} T_{\rm{CMB}}$ is the observed dimensionless photon frequency.  We derived analytic expressions for these functions.  The present formulas are applicable to non-thermal electron distributions as well as the standard thermal distribution.  With these formulas, five-dimensional integrals for solving the Boltzmann equations were reduced to one-dimensional integral of the electron velocity $\beta$.

  In Section 3, we derived analytic expressions for $F_{0}(\beta,x)$, $G_{\ell}(\beta,x)$ and $H_{\ell}(\beta,x)$ in the power series expansion approximation of $\beta$.  These formulas are useful for the calculation in the $\beta \ll 1$ region, where numerical errors of the analytic expressions become non-negligible.  It was also found that the Boltzmann equation in the Fokker$-$Planck approximation was derived with these approximate formulas.

  In Section 4, we calculated the integrated photon redistribution functions which appeared in the Boltzmann equations of the CMB and OBS frames.  

  In Section 5, we studied numerically the analytic formulas and the power series expansion approximation of the integrated photon distribution functions for their applicable $(\beta,x)$ region.  By combining two formulas, we offered a simple and accurate tool to analyse observation data.

  Finally, we derived analytic expressions for the photon redistribution functions $P_{\ell}(s,\beta)$ in Appendix A.  The present formulas are equivalent to those obtained in \cite{noza13}.

\section*{Acknowledgements}

We thank our referee for valuable suggestions.

\appendix

\section{Analytic expressions for the photon redistribution functions}

  We study analytic expressions for the photon redistribution functions $P_{\ell}(s,\beta)$.  The explicit form for $P_{0}(s,\beta)$ was derived by \cite{farg97} and \cite{sazo00}.  The explicit forms for $P_{\ell}(s,\beta)$ were also given in equations (62)--(67) of \cite{noza13}.  In this paper, we derive alternative (but equivalent) expressions which are more suitable for the present calculation.

  We first introduce the photon redistribution functions $P_{\ell,p}(s,\beta)$ and $P_{\ell,n}(s,\beta)$ as follows:
\begin{eqnarray}
P_{\ell}(s,\beta) = \left\{
\begin{array}{ll}
P_{\ell,p}(s,\beta)  &\quad  {\rm for} \, \, \, s > 0 \\
   \\
P_{\ell,n}(s,\beta)  &\quad  {\rm for} \, \, \, s < 0 \\
\end{array}
\right.  \, .
\label{eqA-1}
\end{eqnarray}
Rewriting hyperbolic functions in equations (62)--(67) of \cite{noza13} to functions of ${\rm{e}}^{\pm s}$, one finally obtains the following expressions:
\begin{eqnarray}
P_{0,p}(s,\beta) = \alpha \bigg[ \sum_{n=0}^{3} \, a_{n}(\beta) \, {\rm{e}}^{n s}  
\nonumber  \\
&&\hspace{-25mm}
 + \, a_{4}(\beta) \left( \lambda_{\beta}-s \right) \left( {\rm{e}}^{s}+{\rm{e}}^{2s} \right) \bigg]  \, ,
\label{eqA-2}  \\
&& \hspace{-48mm}
P_{0,n}(s,\beta) = \alpha \bigg[ \sum_{n=0}^{3} \, a_{3-n}(\beta) \, {\rm{e}}^{n s}  \nonumber  \\
&& \hspace{-25mm}
 + \, a_{4}(\beta) \left( \lambda_{\beta}+s \right) \left( {\rm{e}}^{s}+{\rm{e}}^{2s} \right) \bigg]  \, , 
\label{eqA-3}
\end{eqnarray}
\begin{eqnarray}
P_{1,p}(s,\beta) = \alpha \bigg[ \, \sum_{n=0}^{3} b_{n}(\beta) \, {\rm{e}}^{n s} + \, b_{4}(\beta) \left( \lambda_{\beta}-s \right) \left( 1+{\rm{e}}^{3s} \right)   \nonumber  \\
&& \hspace{-62mm}
+ \, b_{5}(\beta) \left( \lambda_{\beta}-s \right) \left( {\rm{e}}^{s}+{\rm{e}}^{2s} \right)  \bigg]  \, ,
\label{eqA-4}  \\
&& \hspace{-86mm}
P_{1,n}(s,\beta) = \alpha \bigg[ \, \sum_{n=0}^{3} b_{3-n}(\beta) \, {\rm{e}}^{n s}  + \, b_{4}(\beta) \left( \lambda_{\beta}+s \right) \left( 1+{\rm{e}}^{3s} \right)   \nonumber  \\
&& \hspace{-62mm}
+ \, b_{5}(\beta) \left( \lambda_{\beta}+s \right) \left( {\rm{e}}^{s}+{\rm{e}}^{2s} \right)  \bigg]  \, ,
\label{eqA-5}
\end{eqnarray}
\begin{eqnarray}
P_{2,p}(s,\beta) = \alpha \bigg[ \sum_{n=-1}^{4} c_{n}(\beta) \, {\rm{e}}^{n s}  + \, c_{5}(\beta) \left( \lambda_{\beta}-s \right) \left( 1+{\rm{e}}^{3s} \right)   \nonumber  \\
&& \hspace{-63mm}
+ \, c_{6}(\beta) \left( \lambda_{\beta}-s \right) \left( {\rm{e}}^{s}+{\rm{e}}^{2s} \right)  \bigg]  \, ,
\label{eqA-6}  \\
&& \hspace{-87mm}
P_{2,n}(s,\beta) = \alpha \bigg[ \sum_{n=-1}^{4} c_{3-n}(\beta) \, {\rm{e}}^{n s} + \, c_{5}(\beta) \left( \lambda_{\beta}+s \right) \left( 1+{\rm{e}}^{3s} \right)   \nonumber  \\
&& \hspace{-63mm}
+ \, c_{6}(\beta) \left( \lambda_{\beta}+s \right) \left( {\rm{e}}^{s}+{\rm{e}}^{2s} \right)  \bigg]  \, ,
\label{eqA-7}
\end{eqnarray}
where $\alpha = 3/(32\beta^{2} \gamma^{2})$.  Note that equation (\ref{eqA-1}) satisfies the relation
\begin{eqnarray}
P_{\ell}(-s,\beta) = {\rm{e}}^{-3 s} \, P_{\ell}(s,\beta) \, .
\label{eqA-8}
\end{eqnarray}

  The coefficients $a_{n}(\beta)$, $b_{n}(\beta)$ and $c_{n}(\beta)$ are given by
\begin{eqnarray}
a_{0}(\beta) = + \frac{1}{\beta^{4} \gamma^{4}}  \, ,
\label{eqA-9}  \\
&& \hspace{-31mm}
a_{1}(\beta) = + \frac{1}{\beta^{4}} \left(1+\beta\right) \left(9 + 3\beta -13\beta^{2} + \beta^{3} + 4\beta^{4}\right)  \, ,
\label{eqA-10}  \\
&& \hspace{-31mm}
a_{2}(\beta) = - \frac{1}{\beta^{4}} \left(1-\beta\right) \left(9 - 3\beta -13\beta^{2} - \beta^{3} + 4\beta^{4}\right)  \, ,
\label{eqA-11}  \\
&& \hspace{-31mm}
a_{3}(\beta) = - \frac{1}{\beta^{4} \gamma^{4}}  \, ,
\label{eqA-12}  \\
&& \hspace{-31mm}
a_{4}(\beta) = - \frac{2}{\beta^{4} \gamma^{2}} \left(3 - \beta^{2}\right) \, ,
\label{eqA-13}
\end{eqnarray}
\begin{eqnarray}
b_{0}(\beta) = + \frac{1}{3 \beta^{6}} \left(1+\beta\right)^{2} \left(55 - 80\beta - 29\beta^{2} + 70\beta^{3} \right.  \nonumber  \\
&& \hspace{-25mm}
\left. - 10\beta^{4} - 4\beta^{5} \right)  \, ,
\label{eqA-14}  \\
&& \hspace{-76mm}
b_{1}(\beta) = + \frac{1}{\beta^{6} \gamma^{2}} \left(45 + 90\beta - 45\beta^{2} - 78\beta^{3} + 10\beta^{4} \right.  \nonumber  \\
&& \hspace{-14mm}
\left.  + 8\beta^{5} \right)  \, ,
\label{eqA-15}  \\
&& \hspace{-76mm}
b_{2}(\beta) =- \frac{1}{\beta^{6} \gamma^{2}} \left(45 - 90\beta - 45\beta^{2} + 78\beta^{3} + 10\beta^{4} \right.  \nonumber  \\
&& \hspace{-14mm}
\left.  - 8\beta^{5} \right)  \, ,
\label{eqA-16}  \\
&& \hspace{-76mm}
b_{3}(\beta) = - \frac{1}{3 \beta^{6}} \left(1-\beta\right)^{2} \left(55 + 80\beta - 29\beta^{2} - 70\beta^{3} \right.  \nonumber  \\
&& \hspace{-25mm}
\left. - 10\beta^{4} + 4\beta^{5} \right)   \, ,
\label{eqA-17}  \\
&& \hspace{-76mm}
b_{4}(\beta) = - \frac{1}{\beta^{6} \gamma^{4}} \left(5 - 3\beta^{2}\right)  \, ,
\label{eqA-18}  \\
&& \hspace{-76mm}
b_{5}(\beta) = - \frac{1}{\beta^{6} \gamma^{2}} \left(45 - 54\beta^{2} + 13 \beta^{4}\right)  \, ,
\label{eqA-19}
\end{eqnarray}
\begin{eqnarray}
c_{-1}(\beta) = + \frac{1}{40 \beta^{8}} \left(1+\beta\right)^{3} \left(105 - 315\beta + 240\beta^{2}  \right.  \nonumber  \\
&& \hspace{-49mm}
\left. + 120\beta^{3} - 213\beta^{4} + 39\beta^{5} + 32\beta^{6}\right)  \, ,
\label{eqA-20}  \\
&& \hspace{-74mm}
c_{0}(\beta) = + \frac{1}{8 \beta^{8} \gamma^{4}} \left(1225 + 840\beta - 1220\beta^{2} - 680\beta^{3}  \right.  \nonumber  \\
&& \hspace{-26mm}
\left.  + 215\beta^{4} + 64\beta^{5}\right)   \, ,
\label{eqA-21}  \\
&& \hspace{-74mm}
c_{1}(\beta) = + \frac{1}{2 \beta^{8} \gamma^{2}} \left(525 + 1260\beta - 885\beta^{2} - 1920\beta^{3}  \right.  \nonumber  \\
&& \hspace{-49mm}
\left.  + 435\beta^{4} + 756\beta^{5}  - 55\beta^{6} - 48\beta^{7}\right)  \, ,
\label{eqA-22}  \\
&& \hspace{-74mm}
c_{2}(\beta) = - \frac{1}{2 \beta^{8} \gamma^{2}} \left(525 - 1260\beta - 885\beta^{2} + 1920\beta^{3}  \right.  \nonumber  \\
&& \hspace{-49mm}
\left.  + 435\beta^{4} - 756\beta^{5}  - 55\beta^{6} + 48\beta^{7}\right)  \, ,
\label{eqA-23}  \\
&& \hspace{-74mm}
c_{3}(\beta) = -\frac{1}{8 \beta^{8} \gamma^{4}} \left(1225 - 840\beta - 1220\beta^{2} + 680\beta^{3}  \right.  \nonumber  \\
&& \hspace{-26mm}
\left.  + 215\beta^{4} - 64\beta^{5}\right)   \, ,
\label{eqA-24}  \\
&& \hspace{-74mm}
c_{4}(\beta) = -\frac{1}{40 \beta^{8}} \left(1-\beta\right)^{3} \left(105 + 315\beta + 240\beta^{2}  \right.  \nonumber  \\
&& \hspace{-49mm}
\left. - 120\beta^{3} - 213\beta^{4} - 39\beta^{5} + 32\beta^{6}\right)   \, ,
\label{eqA-25}  \\
&& \hspace{-74mm}
c_{5}(\beta) = -\frac{3}{2 \beta^{8} \gamma^{4}} \left(35 - 40\beta^{2} + 9\beta^{4}\right)  \, ,
\label{eqA-26}  \\
&& \hspace{-74mm}
c_{6}(\beta) = - \frac{1}{\beta^{8} \gamma^{2}} \left(315 - 585\beta^{2} + 321\beta^{4} - 47\beta^{6}\right)  \, .
\label{eqA-27}
\end{eqnarray}
With these coefficients, it is straightforward to show that equations (\ref{eqA-2})--(\ref{eqA-7}) satisfy a useful relation
\begin{eqnarray}
P_{\ell,n}(s,\beta) = - P_{\ell,p}(s,-\beta)  \, \, \, \,  {\rm for} \, \, \, s < 0 \, ,
\label{eqA-28}
\end{eqnarray}
which simplifies the calculation in Section 2.  In deriving equation (\ref{eqA-28}), we extended $P_{\ell,p}(s,\beta)$ to $s < 0$ region, because they are continuous functions for $-\infty < s < \infty$.

  Finally, we note the following useful relations for the functions $P_{\ell,p}(s,\beta)$ and $P_{\ell,n}(s,\beta)$ at specific values of $s$:
\begin{eqnarray}
P_{\ell,p}(0,\beta) = P_{\ell,n}(0,\beta)  \, ,
\label{eqA-29}  \\
&& \hspace{-40mm}
P_{\ell,p}(\lambda_{\beta},\beta) = P_{\ell,n}(-\lambda_{\beta},\beta) = 0 \, .
\label{eqA-30}
\end{eqnarray}
These relations will be used in the derivation of equations (\ref{eq2-2-5}) and (\ref{eq2-2-13}) in Section 2.  For the derivatives $P_{\ell,p}^{\prime}(s,\beta)$ and $P_{\ell,n}^{\prime}(s,\beta)$ in terms of $s$, one has as follows:
\begin{eqnarray}
P_{\ell,p}^{\prime}(0,\beta) - P_{\ell,n}^{\prime}(0,\beta) =  - \frac{3}{4\beta^{2}} (1-\beta^{2})  \, ,
\label{eqA-31}  \\
&& \hspace{-64mm}
P_{\ell,p}^{\prime}(\lambda_{\beta},\beta) = (-1)^{\ell+1} \frac{3}{8\beta^{2}}(1+\beta)^{3}  \, ,
\label{eqA-32}  \\
&& \hspace{-64mm}
P_{\ell,n}^{\prime}(-\lambda_{\beta},\beta) = (-1)^{\ell} \frac{3}{8\beta^{2}}(1-\beta)^{3}  \, .
\label{eqA-33}
\end{eqnarray}
These relations will be used in the derivation of equation (\ref{eq2-2-13}) in Section 2.

\label{lastpage}


\begin{thebibliography}{99}
\bibitem[\protect\citeauthoryear{Allen, Schmidt \& Fabian}{2002}]{alle02} Allen S. W. , Schmidt R. W., Fabian A. C., 2002, MNRAS, 335, 256
\bibitem[\protect\citeauthoryear{Birkinshaw}{1999}]{birk99} Birkinshaw M., 1999, Phys. Rep., 310, 97
\bibitem[\protect\citeauthoryear{Carlstrom, Holder \& Reese}{2002}]{carl02} Carlstrom J. E., Holder G. P., Reese E. D., 2002, ARA\&A, 40, 643
\bibitem[\protect\citeauthoryear{Challinor \& Lasenby}{1998}]{chal98} Challinor A., Lasenby A., 1998, ApJ, 499, 1
\bibitem[\protect\citeauthoryear{Challinor \& Lasenby}{1999}]{chal99} Challinor A., Lasenby A., 1999, ApJ, 510, 930
\bibitem[\protect\citeauthoryear{Chluba, H\"utsi \& Sunyaev}{2005}]{chlu05} Chluba J., H\"utsi G., Sunyaev R. A., 2005, A\&A, 434, 811
\bibitem[\protect\citeauthoryear{Chluba et al.}{2012}]{chlu12} Chluba J., Nagai D., Saznov S., Nelson K., 2012, MNRAS, 426, 510
\bibitem[\protect\citeauthoryear{Fargion, Konoplich \& Salis}{1997}]{farg97} Fargion D., Konoplich R. V., Salis A., 1997, Z. Phys. C. 74, 571
\bibitem[\protect\citeauthoryear{Hansen}{2004}]{hans04} Hansen S. H., 2004, MNRAS, 351, L5
\bibitem[\protect\citeauthoryear{Itoh, Kohyama \& Nozawa}{1998}]{itoh98} Itoh N., Kohyama Y., Nozawa S., 1998, ApJ, 502, 7
\bibitem[\protect\citeauthoryear{Itoh, Nozawa \& Kohyama}{2000}]{itoh00} Itoh N., Nozawa S., Kohyama Y., 2000, ApJ, 533, 588
\bibitem[\protect\citeauthoryear{Kompaneets}{1956}]{komp56} Kompaneets A. S., 1956, Sov. Phys.$-$JETP, 31, 876
\bibitem[\protect\citeauthoryear{Nozawa, Itoh \& Kohyama}{1998}]{noza98} Nozawa S., Itoh N., Kohyama Y., 1998, ApJ, 508, 17
\bibitem[\protect\citeauthoryear{Nozawa, Itoh \& Kohyama}{2005}]{noza05} Nozawa S., Itoh N., Kohyama Y., 2005, A\&A, 440, 39
\bibitem[\protect\citeauthoryear{Nozawa \& Kohyama}{2013}]{noza13} Nozawa S., Kohyama Y., 2013, MNRAS, 434, 710
\bibitem[\protect\citeauthoryear{Planck Collaboration}{2011}]{plan11} Planck Collaboration et al., 2011, A\&A, 536, A8
\bibitem[\protect\citeauthoryear{Rephaeli}{1995}]{reph95} Rephaeli Y., 1995, ApJ, 445, 33
\bibitem[\protect\citeauthoryear{Sazonov \& Sunyaev}{1998}]{sazo98} Sazonov S. Y., Sunyaev R. A., 1998, Astron. Lett. 24, 553
\bibitem[\protect\citeauthoryear{Sazonov \& Sunyaev}{2000}]{sazo00} Sazonov S. Y., Sunyaev R. A., 2000, ApJ, 543, 28
\bibitem[\protect\citeauthoryear{Sunyaev \& Zeldovich}{1980}]{suny80} Sunyaev R. A., Zeldovich Ya. B., 1980, MNRAS, 190, 413
\bibitem[\protect\citeauthoryear{Sunyaev \& Zeldovich}{1981}]{suny81} Sunyaev R. A., Zeldovich Ya. B., 1981, Astrophys. Space Phys. Rev., 1, 1
\bibitem[\protect\citeauthoryear{Wright}{1979}]{wrig79} Wright E. L., 1979, ApJ, 232, 348
\bibitem[\protect\citeauthoryear{Zeldovich \& Sunyaev}{1969}]{zeld69} Zeldovich Y. B., Sunyaev R. A., 1969, Ap\&SS, 4, 301
\bibitem[\protect\citeauthoryear{Zemcov et al.}{2010}]{zemc10} Zemcov M. et al., 2010, A\&A, 518, L16
\bibitem[\protect\citeauthoryear{Zdziarski \& Pjanka}{2013}]{zdzi13} Zdziarski A. A., Pjanka P., 2013, MNRAS, 436, 2950
\end{thebibliography}
\end{document}